\documentclass[aps,prl,twocolumn,showpacs,superscriptaddress]{revtex4}
\def\frac#1#2{{\textstyle{#1\over#2}}} 

\def\ket#1{| #1\rangle}

\def\R{\hbox{\rm I \kern-5pt R}}

\def\ajou#1&#2(#3){\ \sl#1\bf#2\rm(19#3)}

\usepackage{amssymb}
\usepackage{amsmath}
\usepackage[sort&compress]{natbib}
\usepackage[dvips]{graphicx}

\begin{document}

\title{No Signalling and Quantum Key Distribution} 

\author{Jonathan Barrett}
\email{jbarrett@perimeterinstitute.ca} 
\affiliation{Physique Th\'{e}orique, Universit\'{e}
  Libre de Bruxelles, CP 225, Boulevard du Triomphe, 1050 Bruxelles, Belgium}
\affiliation{Centre for Quantum Information and Communication, CP 165/59,
  Universit\'{e} Libre de Bruxelles, Avenue F. D. Roosevelt 50, 1050 Bruxelles,
  Belgium} 

\author{Lucien Hardy} \email{lhardy@perimeterinstitute.ca}
\affiliation{Perimeter Institute, 35 King Street North, Waterloo ON, N2J 2W9,
  Canada} 

\author{Adrian Kent} \email{A.P.A.Kent@damtp.cam.ac.uk}
\affiliation{Centre for Quantum Computation, DAMTP, Centre for Mathematical 
 Sciences, \\ University of Cambridge,
  Wilberforce Road, Cambridge CB3 0WA, U.K.}

\date{March 2005 (revised)} 

\begin{abstract}
  Standard quantum key distribution protocols are provably secure
  against eavesdropping attacks, if quantum theory is correct.  It is
  theoretically interesting to know if we need to assume the validity
  of quantum theory to prove the security of quantum key distribution,
  or whether its security can be based on other physical principles.
  The question would also be of practical interest if quantum
  mechanics were ever to fail in some regime, because a scientifically
  and technologically advanced eavesdropper could perhaps use
  post-quantum physics to extract information from quantum
  communications without necessarily causing the quantum state
  disturbances on which existing security proofs rely.  Here we
  describe a key distribution scheme provably secure against general
  attacks by a post-quantum eavesdropper who is limited only by the
  impossibility of superluminal signalling. The security of the scheme
  stems from violation of a Bell inequality.
\end{abstract}


\maketitle

With the discoveries of quantum cryptography \cite{wiesner} and
quantum key distribution \cite{BBef,ekert}, it is now well understood that
cryptographic tasks can be guaranteed secure by physical principles.  For
example, we now have protocols for various important tasks, including key
distribution, that are provably secure provided quantum 
theory is correct \cite{list}.  
Protocols for bit commitment have been
developed with security based only on the impossibility of superluminal
signalling \cite{akrel,akrelfinite}. The possibility of basing cryptographic
security on known superselection rules has also recently been 
discussed \cite{vc,kmp}.

In this paper we investigate whether it is possible to devise a quantum key
distribution scheme that is provably secure if superluminal signalling is
impossible.  We allow for
eavesdroppers who can break the laws of quantum mechanics, as long as nothing
they can do implies the possibility of superluminal signalling. In general,
this will mean that the security proofs of existing quantum key distribution
protocols are no longer valid, as we can no longer assume that quantum theory
correctly predicts the tradeoff between the information that Eve can extract
and the disturbance she must necessarily cause.

As we show below, there is an intimate connection
between the possibility of such a protocol and the violation of a Bell
inequality \cite{bell,chsh}.  Non-local (in the sense of Bell inequality violating)
correlations constitute an exploitable resource for this task, just as
entanglement is a resource for conventional quantum key distribution.
We present a quantum scheme, involving Bell violation, that is secure
against general attacks by a non-signalling Eve.  

One motivation for this work is practical: existing
security proofs assume the validity of quantum theory, and while
quantum theory has been confirmed in an impressive range of experiments, it
remains plausible that some future experiment will demonstrate a limit to its
domain of validity.  Admittedly, it is also conceivable that some future
experiment could demonstrate the possibility of superluminal signalling.  But
the possibilities are logically independent: quantum theory could fail without
violating standard relativistic causality, and vice versa.  A cryptographic
scheme that can be guaranteed secure by either of two physical principles is
more trustworthy than one whose security relies entirely on one. 

There are also compelling theoretical motivations. 
Understanding which cryptographic tasks can be guaranteed secure by
which physical principles improves our understanding of the
relationship between information theory and physical theory. 
Our work also demonstrates a new way of proving security for
quantum protocols, which may be useful in other contexts,  
and sheds new light on non-locality and its
relation to secrecy. 

\subsection{A Quantum Protocol for Secret Bit Distribution}

We assume that Alice and Bob have a noise-free quantum
channel and an authenticated classical channel.  
Consider the following protocol, which we show below
generates a single shared secret bit, guaranteed secure
against general attacks by post-quantum eavesdroppers.  Define
the bases $X_r = \{ \cos \frac{r \pi}{2 N} \ket{0} + 
\sin \frac{r \pi}{2 N} \ket{1}, 
- \sin \frac{r \pi}{2 N} \ket{0} + 
\cos \frac{r \pi}{2 N} \ket{1} \}$ for integer $r$. 
For each basis, we define outcomes $0$ and $1$ to correspond 
respectively to the projections onto the first and second basis elements.
Thus $X_{r+N}$ contains the same basis states as $X_r$ with the 
outcome conventions
reversed; i.e. we interpret the bases $X_{-1}$ and $X_N$ below
to be $X_{N-1}$ and $X_0$ with outcomes reversed. 
We take the security parameters $N$ and $M$ (defined below) 
to be large positive integers.  To simplify the analysis,
we will take $M \ll N$.  
\begin{enumerate}
\item Alice and Bob share $n= M N^2$ pairs of systems, each in the maximally
  entangled state $\ket{\psi_{-}}=1/\sqrt{2} (\ket{01}-\ket{10})$.
\item  Alice and Bob choose independent random elements $r^i_A$ and $r^i_B$
of the set $\{ 0, 1, \ldots , N-1 \}$ for each $i$ from $1$ to $M N^2$, 
and measure their $i$-th particle in the bases $A_i \equiv X_{r^i_A}$ and $
B_i \equiv X_{r^i_B}$.   
\item  When all their measurements are complete, Alice and Bob announce their
  bases over a public, authenticated, classical channel.
\item  Alice and Bob abort the protocol and restart unless
$$ 2MN \leq \sum_i \sum_{c=-1,0,1} | \{ j : A_j = X_i , B_j = X_{i+c} \} | \, . $$
(The expected size of the sum is $3MN$.  
The probability of the condition failing is 
of order $e^{-MN/6}$.)   
\item  The outcomes are kept secret for one randomly chosen
pair for which the bases chosen were $X_i$ and $X_{i+c}$ for some
$i$ and $c=-1,0$ or $1$.  We call bases of this form {\it neighbouring
or identical}.  The outcomes are announced
for all the remaining pairs (for all basis choices). 
\item Alice and Bob abort the protocol if their outcomes $a$ and $b$
are not anti-correlated (i.e. $a \neq b$) in all the cases where they
chose neighbouring or identical bases. 
\item If the protocol is not aborted, their unannounced outcomes
define the secret bit, which is taken by Alice to be equal to her outcome
and by Bob to be opposite to his. 

\end{enumerate}

\subsection{Eavesdropping attacks}

To analyse the security of this protocol, we must describe formally
the actions available to post-quantum eavesdroppers. 
To give Eve maximum power, we assume that each pair  
of systems is produced by a source under her control.
In a general, or collective, attack,
Eve prepares $2n+1$ systems in a post-quantum state $\lambda$, sending $n$
systems to Alice, $n$ to Bob, and keeping $1$.  The state $\lambda$
defines measurement probabilities
\[
\mathrm{P}_{\lambda}(a b e| A B E),
\]
where $A= \{ A_1,\ldots,A_n \}$, $B= \{ B_1,\ldots,B_n \}$ are sets
of Alice's and Bob's possible measurement choices and $E = \{ E_1 \}$
is a set containing a possible measurement choice of Eve,
with corresponding outcomes $a,b,e$.
This state may be non-quantum and non-local, but must not allow
signalling even if the parties cooperate.  Thus, for any partitionings
$A= A^1 \cup A^2$, $B= B^1 \cup B^2$ and 
$E = E^1 \cup E^2$ (possibly including empty subsets), 
and any alternative choices 
$\bar{A}^2 , \bar{B}^2 , \bar{E^2}$,
we require that 
\begin{eqnarray}\label{nosignal}
\lefteqn{\sum_{a^2 b^2 e_2} 
P_{\lambda} ( a^1 a^2 b^1 b^2 e_1 e_2 \, | \, 
A^1 A^2 B^1 B^2 E^1 E^2 ) = }\hspace{50pt} \\
\nonumber
&& \sum_{\bar{a}^2 \bar{b}^2 \bar{e}_2} 
P_{\lambda} ( a^1 \bar{a}^2 b^1 \bar{b}^2 
e_1 \bar{e}_2 \, | \, 
A^1 \bar{A}^2 B^1 \bar{B}^2 E^1 \bar{E}^2 ) \, .
\end{eqnarray}
Eve may wait until 
all Alice's and Bob's communications are finished before performing her
measurement.

We need a further technical assumption.  
It seems natural to postulate
that, once Eve has prepared a post-quantum state
$\lambda$, the range of measurements available to her and their
outcome probabilities are (up to relabellings) \emph{time-independent}.  
In fact, a slightly weaker assumption suffices: we assume that
in post-quantum theory, as in quantum theory, measurements on 
a shared state cannot be used to send signals between the parties
in any configuration (even if not spacelike separated). 
If this assumption were dropped, one could allow a theory in which information 
about the bases and outcomes of any measurements carried out by Alice and Bob 
propagates to Eve at light speed, so she can 
obtain these data by a later measurement
timelike separated from Alice's and Bob's.  
While theories of this type may seem implausible, or even pathological, 
they can be made internally consistent without allowing 
superluminal communication \cite{aknonlinear}.  
Clearly, secure key distribution would be impossible if Eve 
could exploit a theory of this type. 

One can justify excluding this possibility by extending a standard
cryptographic assumption to post-quantum cryptology.  
Conventional security analyses of quantum
key distribution require that Alice's and Bob's laboratories 
are completely secure against Eve's scrutiny --- 
a necessary cryptographic assumption, which does not follow
from the laws of quantum theory.  Similarly, in the post-quantum context,  
we assume that {\it no} information about events in Alice's and Bob's 
laboratories --- in particular, their measurements or outcomes --- 
subsequently propagates to Eve.  Put another way, Alice and Bob have
to assume they can establish secure laboratories, else cryptography
is pointless.  The aim is to guarantee secure key distribution
modulo this assumption.  We shall prove that the protocol above is 
indeed secure against general attacks.

\subsection{Proof of security}

We define $A_j$, $B_j$ to be Alice's and Bob's basis choices for the $j$-th
pair; these are random variables, each measurement occuring with probability
$1/N$. We also define $a_j$, $b_j$ to be their measurement outcomes and write 
\begin{eqnarray*}
\lefteqn{t_j = \frac{1}{3N} \sum_{c=-1,0,1} \sum_{i=0}^{N-1}}\hspace{30pt}\\
&& \mathrm{P}_{\lambda} ( a_j \neq b_j \, | \, A_j = X_i \, , \, B_j = X_{i+c} ). 
\end{eqnarray*}
(Recall that $X_{-1}$ and $X_N$ are $X_{N-1}$ and $X_0$ with outcomes reversed.)
Note that if $\lambda$ is local we have $t_j\leq 1 - \frac{2}{3N}$. Thus this is a
generalised Bell inequality (it is in fact similar to the chained Bell
inequality of Braunstein and Caves \cite{braunsteincaves}.)  If there is no
eavesdropping, so that genuine singlet states are shared, then quantum
mechanics gives $t_j = 1 - O ( 1/(N^2) )$, for all $j$, thus violating the
inequality for large enough $N$. This is crucial for the security of the
protocol; it is violation of this inequality that allows Eve's knowledge to
be bounded. Below, we shall derive a lower bound on the value of $t_s$ for the
secret pair $s$, given that 
Alice's and Bob's tests are passed, and given that Eve
is not using a strategy that almost always fails the tests. Then we show that
the lower bound on $t_s$ implies an upper bound on Eve's information, which
can be made arbitrarily small as $M,N$ become large.

From now on, we assume that there is at least one pair for which Alice's and
Bob's measurements were neighbouring or identical (otherwise they will abort).
Let $s$, a random variable, be the index of the pair chosen to define the
secret bit.  A post-quantum state $\lambda$ determines the probability
$\mathrm{P}_{\lambda} ({\rm pass})$ that Alice's and Bob's tests are
passed, so that they do not abort the protocol.

{\bf Lemma} \qquad 
For any $\lambda$ such that $\mathrm{P}_{\lambda}(\mathrm{pass})>\epsilon$, we have that
\[
\mathrm{P}_{\lambda}( a_s \neq b_s | \mathrm{pass}) > 1 - 1/(2MN\epsilon).
\]

{\bf Proof} \qquad Let $m$, a random variable, be the number of pairs for
which the measurements were neighbouring or identical. For a given pair, let
$C$ be the condition that the measurements were neighbouring or identical and
the outcomes anti-correlated. If the secret pair satisfies $C$, then Alice and
Bob will agree on the value of the secret bit. We denote by $\#(C)$ the number
of pairs for which $C$ holds. Define the following four mutually exclusive and
collectively exhaustive events.
\begin{eqnarray*}
&E_0& \ m < 2 MN\\
&E_1& \ m \geq 2MN \mathrm{\ and\ }\#(C) < m-1\\
&E_2& \ m \geq 2MN \mathrm{\ and\ }\#(C) = m-1\\
&E_3& \ m \geq 2MN \mathrm{\ and\ }\#(C) = m.
\end{eqnarray*}
Note
that if $E_0$ or $E_1$ occurs, then Alice and Bob will definitely abort. If $E_3$ occurs,
then Alice and Bob will definitely not abort. A given post-quantum state $\lambda$ 
defines a probability for each of these four
events, which we write as $\mathrm{P}_{\lambda}(E_i)\equiv q_i$.

Now we have $\mathrm{P}_{\lambda}(\mathrm{pass}) = q_3 + q_2 \, \mathrm{P}_{\lambda}(\mathrm{pass}|E_2)$.
If $E_2$ occurs, then the test will only be passed if the secret pair do 
not satisfy $C$. This means that we have
\begin{eqnarray}
\mathrm{P}_{\lambda}(\mathrm{pass}|E_2)& = & \sum_{i=2MN}^{MN^2}\mathrm{P}(m=i|E_2)/i\nonumber\\
&\leq& 1/(2MN).
\end{eqnarray}
But $\mathrm{P}_{\lambda}(\mathrm{pass}) > \epsilon$, so we can write $q_3 > \epsilon - q_2/(2MN)$.
Therefore
\begin{eqnarray}\label{constraint}
\mathrm{P}_{\lambda}(a_s\neq b_s|\mathrm{pass})&=
&\frac{q_3}{q_3+q_2\,P_{\lambda}(\mathrm{pass}|E_2)}\nonumber\\
&>& 1 - 1/(2MN\epsilon),
\end{eqnarray}
where the inequality follows from the fact that the right hand side of the
first line either equals $1$ or is monotonically increasing with $q_3$. QED.

It follows from the lemma above, the 
no-signalling condition (\ref{nosignal})
and the chain rule for conditional probabilities
that, conditioned on passing the test,
\begin{equation}\label{tbound}
t_s > 1 - 1/(2MN\epsilon). 
\end{equation}
From now on, we assume that the test is passed, and we can consider that
Alice, Bob and Eve share three systems, such that Eq.~(\ref{tbound}) is
satisfied. We now show that the knowledge that Eve can get by performing a
measurement on her system is small.

We do this by contradiction. Thus suppose that with probability $\delta>0$,
Eve gets an outcome $e_0$ such that
\begin{eqnarray*}
\lefteqn{\mathrm{P}_{\lambda} 
( a_s = b, \, b_s = \bar{b} \, | \, A_s = X_k, \, B_s = X_{k+d}, \, e_0)}\hspace{140pt}\\ 
&& > (1/2)( 1 + \delta' ), 
\end{eqnarray*}
for some $k$ and $d=-1,0$ or $1$, where $\delta'>0$ and $b\in\{0,1\}$. Define
\begin{eqnarray*}
p^A_i &\equiv& \mathrm{P}_{\lambda} (a_s = b | A_s = X_i \, , \, e_0 )\\
p^B_i &\equiv& \mathrm{P}_{\lambda} (b_s = \bar{b} | B_s = X_{i} \, , \, e_0),
\end{eqnarray*}
The no-signalling condition (\ref{nosignal}) 
ensures that $p^A_i$ is independent of which
measurement Bob performs, and similarly that $p^B_i$ is independent of which
measurement Alice performs. This enables us to write $p^A_{k}, p^B_{k+d} >
(1/2)( 1 + \delta' )$.  
Now 
\begin{eqnarray*} 
\lefteqn{\mathrm{P}_{\lambda} ( a_s \neq b_s \, | \, 
A_s = X_i, \, B_s = X_{i+c}, \, e_0) = }\hspace{50pt} \\
&& \mathrm{P}_{\lambda} ( a_s = b , b_s = \bar{b} \, | \, 
A_s = X_i, \, B_s = X_{i+c}, \, e_0) \\
&+& \mathrm{P}_{\lambda} ( a_s = \bar{b} , b_s = b \, | \, 
A_s = X_i, \, B_s = X_{i+c}, \, e_0) \\
&\leq & \min ( p_i^A , p_{i+c}^B ) + \min ( 1 - p_i^A , 1 - p_{i+c}^B ) \\
&=& 1 - | p^A_i - p^B_{i+c} | \, . 
\end{eqnarray*}
Now, using (\ref{nosignal}) again and the triangle
inequality, we have
\begin{eqnarray*} 
\lefteqn{\sum_{c= -1,0,1} 
\sum_{i=0}^{N-1} \mathrm{P}_{\lambda} ( a_s \neq b_s \, | \, 
A_s = X_i, \, B_s = X_{i+c}, \, e_0)}\hspace{50pt}\\
&\leq&  3 N - \sum_{c=-1,0,1} \sum_{i=0}^{N-1} | p^A_i - p^B_{i+c} |\\
&\leq& 3N  - \sum_{i=0}^{N-1} | p^A_i - p^A_{i+1} | \\
& \leq &  3N - | 2 p^A_k - 1 | \\
&\leq& 3N - \delta'  \, .
\end{eqnarray*}
This implies that, conditioned only on passing the test,
\begin{equation}\label{tdelta}
t_s \leq 1 -  (\delta \delta')/(3N).
\end{equation}
For any fixed $\delta , \delta' > 0$, we can choose $M, N,\epsilon$ such that
this is inconsistent with Eq.~(\ref{tbound}). $M$ must also be chosen so that
quantum correlations are unlikely to fail the test.  For example, taking $ M =
N^{3/4} \, , \epsilon = N^{-1/4} $ achieves this for sufficiently large $N$.
(Note that if Alice's and Bob's outcomes are classically correlated via a local
hidden variable theory, the chances
of passing the test are very small, and there exists no choice of parameters
for which Eqs.~(\ref{tbound}) and (\ref{tdelta}) are inconsistent.) 

Although we restricted the security parameter $M \ll N$ to simplify 
the discussion, the protocol can be generalised to allow $M$ arbitrarily
large. In this case, Alice's and Bob's security test is that the 
number of pairs for which the outcomes are not anti-correlated
should be statistically consistent with quantum predictions;
the method of our security proof generalises to cover this case.  

\subsection{Discussion}

The above security proof shows that our protocol allows Alice and Bob
to generate a single shared bit and guarantee its security
even against collective attacks by a post-quantum Eve. 
The protocol can be generalised to generate an arbitrary 
shared secret bit string, with the same security guarantee. 

Non-locality is crucial to the success of the protocol. It is easy to see that
if Alice and Bob were violating no Bell inequality, then Eve could eavesdrop
perfectly by preparing each pair of systems in a post-quantum state that is
deterministic (where deterministic means that all probabilities defined by the
state are $0$ or $1$) and local. This would give Eve perfect information about
Alice's and Bob's measurement outcomes.  On the other hand, if Alice and Bob
are violating a Bell inequality, then at least some of the post-quantum states
prepared by Eve must be non-local. But any state that is deterministic and
non-local allows signalling \cite{valentini}. So this trivial eavesdropping
strategy is not available to Eve.

More generally, we can say that the protocol works because, once the
no-signalling condition is assumed, non-local correlations satisfy
a monogamy condition analogous to that of entanglement in quantum
theory. The monogamy of
non-locality was first noted in Ref.~\cite{barrettetal}, where it was shown
that no signalling implies that there exist certain sets of non-quantum
correlations such that Alice's and Bob's outcomes cannot be correlated with a
third party. Here we have shown that there are quantum correlations with the
same property, and used these to construct a key distribution protocol.

It is interesting to contrast the Ekert quantum key distribution
protocol \cite{ekert}, in which a test of the Clauser-Horne-Shimony-Holt
(CHSH) inequality \cite{chsh} is performed. It may appear as if non-locality
is playing a crucial r\^ole here, too. In this case, however, the purpose of
the CHSH inequality test is to verify that the shared states are close to
singlets --- and this is a task that other measurements, not involving
non-locality, can perform equally well \cite{bbm}.

{\bf Acknowledgments} \qquad  We thank Daniel Gottesman for stressing the power 
of collective attacks by a post-quantum Eve and spurring our interest
in producing a general security proof.  
JB thanks Nicolas Cerf, Nicolas Gisin, Serge Massar and Stefano Pironio for
helpful discussions and acknowledges financial support from the
Communaut\'{e} Fran\c{c}aise de Belgique grant ARC 00/05-251, the
IUAP programme of the Belgian government grant V-18, and the EU
project RESQ (IST-2001-37559). JB and LH acknowledge HP Bursaries. AK
acknowledges financial support from the EU project PROSECCO (IST-2001-39227) 
and the Cambridge-MIT Institute.

\end{document}